\documentclass[english,keywords,aps,amsmath,amssymb,twocolumn]{revtex4}
\usepackage[T1]{fontenc}
\usepackage[latin1]{inputenc}
\usepackage{babel}
\usepackage{amssymb}
\usepackage{graphicx}
\usepackage{color}
\usepackage{bm}
\usepackage{longtable}
\usepackage{amsmath} 
\usepackage{amsfonts}
\usepackage{amssymb}
\begin{document}
\title{Macrorealistic inequalities stronger than the standard Leggett-Garg inequalities}
\author{Swati Kumari}
\author{A. K. Pan \footnote{akp@nitp.ac.in}}
\affiliation{National Institute of Technology Patna, Ashok Rajpath, Patna, Bihar 800005, India}
\begin{abstract}
In two-party, two-input and two-output measurement scenario only relevant Bell's inequality is the Clauser-Horne-Shimony-Holt (CHSH) form. They also provide the necessary and sufficient conditions for local realism. Any other form, such as, Clauser-Horne and Wigner forms reduce to the CHSH one.  Standard Leggett-Garg inequalities are often considered to be the temporal analog of CHSH inequalities. But, they do not provide the necessary and sufficient conditions for macrorealism. There is thus scope of formulating new macrorealist inequalities different and stronger than the standard Leggett-Garg inequalities for testing compatibility between the macrorealism and quantum theory. In this paper, we propose three different classes of macrorealistic inequalities in three-time scenario; (i) The ones equivalent to standard Leggett-Garg inequalities in both macrorealist model and in quantum theory. (ii) A class of inequalities which are equivalent to the standard ones in macrorealist model but inequivalent and stronger in quantum theory (iii) Another class of inequalities which are inquivalent to the all formulations of Leggett-Garg inequalities both in macrorealist model and in quantum theory. This class of macrorealist inequalities reveals the incompatibility between macrorealism and quantum theory for specific cases even when any other formulation of Leggett-Garg inequalities fail to do so. We extend the formulations of the class (ii) inequalities to the four-time and two-time measurement scenario. Further, we provide a brief discussion about the alternate formulation of macrorealism which was derived based on the no-signaling in time conditions.
\end{abstract}
\maketitle
\section{Introduction}
Bell's inequalities \cite{bell} are at the heart of quantum foundations research, are derived based on the assumption of local realism. The notion of local realism demands that the probability of measurement outcomes are determined by a suitable set of ontic states and are un-influenced by space-like separated measurements. The quantum violation of a Bell's inequalities in quantum theory implies that quantum statistics cannot be reproduced by a local realistic theory.  In his famous work, Fine \cite{fine} argued that for two-party, two-input and two-output Bell scenario (henceforth, $2222$ scenario) only relevant inequality is the  Clauser-Horne-Shimony-Holt(CHSH) one \cite{chsh}. Any other form of inequalities, such as Clauser-Horne \cite{ch} and Wigner form \cite{ep} in the aforementioned scenario reduce to the CHSH one.  Fine's theorem also states that CHSH inequalities provide the necessary and sufficient condition for local realism and certify the existence of the joint probabilities. However, for more than two-measurement scenario, the inequalities inequivalent to CHSH form can be found \cite{col,renou}. 

In $1985$, Leggett and Garg proposed an interesting set of inequalities for testing the status of macrorealism in quantum theory. The concept of macrorealism consists of two main assumptions \cite{leggett85,leggett,A.leggett} which seem reasonable in our everyday world are the following; a) \emph{ Macrorealism per se :} If a macroscopic system has two or more macroscopically distinguishable ontic states available to it, then the system remains in one of those states at all instant of time. b) \emph{Noninvasive measurability :} The definite ontic state of the macrosystem is determined without affecting the state itself or its possible subsequent dynamics.

Thus the violation of standard Leggett-Garg inequalities(LGIs) imply the untenability of either or both the assumptions of \textit{macrorealism per se} and \textit{noninvasive measurability}. In recent times, a flurry of theoritical \cite{bruk,kofler,UshaDevi,budroni,maroney,budroni15,clemente,saha,moreira,clemente16,hall,swati17,hall17,pan17,kumari,pan18,hall19,hall19a} and experimental \cite{a.p,goggin,xu,dressel,suzuki,arndt,julsgaard,gerlich,isart,souza,mahesh11,katiyar13,formaggio,katiyar} works in this issue have been reported in this topic.  However, it remains a debatable issue that how noninvasive measurability can be guaranteed in a real experiment \cite{maroney,hall,hall17,hall19}. 
	
There is a common perception that LGIs are \emph{temporal analog} of CHSH inequalities. This inference is motivated from the structural resemblance between CHSH inequalities and four-time standard LGIs. However, standard LGIs do not  provide the necessary and sufficient  condition (NSC) for macrorealism in contrast to the CHSH inequalities. This simply means that when a LGI is violated by quantum theory, one concludes that the macrorealism is violated, but if LGIs are satisfied, no conclusive argument can be made. In other words, standard LGIs are not a good indicator for capturing the notion of macrorealism. An alternate formulation of macrorealism - the no-signalling in time (NSIT) conditions  is proposed by Clemente and Kofler \cite{clemente,clemente16}. They argued that a suitable combinations of NSIT conditions provide the NSC for macrorealism.  In the macrorealism polytope, standard LGIs do not represent the facets of that polytope, rather it is a hyperplane. We have recently shown \cite{pan17} that there is no connection between the violation of standard LGIs and joint measurability as the violation of standard LGIs can be obtained for any degree of unsharpness of the measurement. We note here that there is another set of LGIs known as Wigner form of LGIs \cite {swati17,pan17} which are stronger than standard LGIs. However, they also do not provide NSC for macrorealism.

A crucial point to note here that the argument of NSIT formulation of macrorealism \cite{clemente16,clemente}   can be considered as a logical proof of macrorealism. This is due to the fact that, in contrast to LGIs, in order a set of NSIT conditions, one has to test a set of equalities. Such a test is nearly impossible to achieve in real experiments. This is quite similar to the Bell's original proof of non-locality \cite{bell} which requires perfect anti-correlation and the logical proof of Kochen-Specker  theorem \cite{ks,ker} that requires deterministic outcome of projective measurements. Both the Bell and the Kochen-Specker theorm are tested through suitably formulated inequalities. Thus, inequality based formulation of macrorealism is needed for  testing the macrorealism experimentally. This simply sets the motivation of the present paper. However, how the NIM condition can be ensured in an experiment is a separate issue. 

In this paper, we formulate several new and interesting classes of macrorealistic inequalities for three-time measurement scenario. (i) The inequalities equivalent to standard LGIs in both macrorealist model and in quantum theory. In particular, we propose a probabilistic form of LGIs involving only pair-wise anti-correlated probabilities and demonstrate the equivalence of them with standard LGIs in both macrorealist model and in quantum theory. (ii) A class of inequalities which are equivalent to the standard ones in macrorealist model but inequivalent and stronger in quantum theory. Wigner form of LGIs falls into this class which are already shown to be stronger than standard ones by us \cite{pan17} by using a different line of argument than the elegant approach adopted here. Further, we formulate a new set of inequalities which we term as Clauser-Horne form of LGIs . We show that this form of LGIs are also stronger than the standard LGIs and in some specific cases stronger than Wigner form of LGIs in quantum theory. We extend the formulation of this class of inequalities to four-time LGIs and compared them with the CHSH scenario.  We also propose the Wigner form of LGIs for two-time measurement scenario and demonstrate the inequivalence with them with the standard two-time LGIs in quantum theory.  (iii) Another class of inequalities is derived which is inquivalent to all the aforementioned LGIs both in macrorealist model and in quantum theory. Importantly, this class of inequalities reveals the incompatibility between macrorealism and quantum theory for specific situations when standard, Wigner and Clauser-Horne formulations of LGIs fail to do so. 

In order to demonstrate our results, by following \cite{hall14}, we develop an approach  to first provide an alternate derivation of CHSH inequalities from the assumption of the existence of triple-wise joint probability distributions. Using that approach, we explicitly show that any formulation of local realistic inequalities are equivalent to the CHSH one. This is already proved by Fine \cite{fine}, and hence not new. But, the aforementioned approach is simple and elegant which enables us to examine the (in)equivalence between various classes of macrorealistic inequalities  and standard LGIs. The above feature of inequivalence is clearly in contrast to the $2222$ Bell scenario where only relevant inequality is the CHSH one. Note that, the measurement schemes involved in Bell and in LG scenarios are very much different and the structural resemblance between CHSH and four-time LGIs is in fact cosmetic.  The statistical version of locality condition in a realist model, i.e., the no-signaling in space condition in an operational theory is always satisfied. In other words, measurement performed in one site cannot disturb the outcome in the other space-like separated site. In LG scenario, the sequential measurement of non-commuting observables is performed. Although joint probability distribution between two non-commuting observables does not exist in quantum theory but the sequential measurement provides a way to compute the sequential joint probabilities for the non-commuting observables by taking into account the disturbance caused by the prior measurement to the future measurements. Due to such disturbance, the statistical version of non-invasive measurability, i.e., no-signaling in time condition is not satisfied in quantum theory, in general. This, in fact, is the root cause of obtaining the inequivalent LGIs in quantum theory, even when they are equivalent to the standard LGIs in a macrorealist model.

This paper is organized as follows. In the Sec.II, we provide an alternate derivation of CHSH  inequalities and show the equivalence between various formulations of local realistic inequalities. Following the approach developed in Sec.II, we propose two classes of of LGIs for three-time measurement scenario in Sec.III, and show that some of the forms are equivalent to the standard LGIs in macrorealist theory as well as in quantum theory. While other forms of LGIs viz., Wigner and Clauser-Horne forms are inequivalent and stronger than the standard LGIs in quantum theory. In Sec.IV, we demonstrate a similar ineqivalence for four-time measurement scenario. In Sec.V, we examine probabilistic form of two-time LGIs, which are shown to be stronger than standard two-time LGIs. We formulate an interesting class of macrorealistic inequalities in Sec.VI which captures the incompatibility between macrorealism and quantum theory in the parameter ranges when no other LGIs do the same. In Sec. VII, we discuss the no-signaling in time formulation of macrorealism. We summarize and discuss our results in the Sec.VIII.
\section{An alternate derivation of CHSH inequalities and Fine's theorem}
 We first provide an alternate derivation of CHSH inequalities by considering four triple-wise joint probabilities. The CHSH scenario involves two space-like separated observers, Alice and Bob, who share a physical system consisting of two subsystems in their possessions and collect statistics to calculate the joint probabilities. Let, Alice and Bob perform  measurement of dichotomic observables $A_{1}, A_{2}$  and  $B_{1}, B_{2}$ on their respective sites, which produce outcomes $a_{1}, a_{2}=\pm1$ and  $b_{1}, b_{2}=\pm1$ respectively.

 The Fine theorem states that the following statements are equivalent. 
	
(i) There exists a global joint probability distribution $P(a_{1}, a_{2},b_{1}, b_{2})$ for all outcomes whose marginals are the experimentally observed probabilities. (ii) There exists a local realistic model for all probabilities. (iii) CHSH inequalities hold.

Busch \cite{busch} provided the derivation of CHSH inequalities assuming the existence of quadruple-wise probability distributions and showed the connection with pair-wise joint measurability of local observables. Anderson \textit{et al.} \cite{ander} derived them assuming the existence of triple-wise joint probability distributions and no-signalling condition.
 Recently, using an interesting approach, Halliwell \cite{hall14} derived  CHSH inequalities using the existence of quadruple-wise probability distribution, subjected to some auxiliary restrictions.  Using Bayes' theorem, it can be written \cite{wolf} that $P(a_{1},a_{2},b_{1},b_{2})=P(a_{1},a_{2},b_{1})P(a_{1},a_{2},b_{2})/P(a_{1},a_{2})$. Here we first re-derive CHSH inequalities from triple-wise joint probability distributions $P(a_{1},a_{2},b_{1})$ and $P(a_{1},a_{2},b_{2})$. Such derivation of CHSH inequality is not new but the way the equivalence between various forms of inequalities in 2222 Bell scenario demonstrated here is interesting. The approach developed here will be used to examine the equivalence between various formulations of LGIs. 

The existence of joint probability $P(a_{1},a_{2},b_{1})$ for the measurement $A_{1}$, $A_{2}$ on Alice side and $B_{1}$ on Bob side provides the pairwise marginals, for example, $P(a_{1},b_{1} )= \sum_{a_{2}=\pm1}{P(a_{1},a_{2},b_{1})}$. Using the moment expansion adopted in \cite{hall14}, we can write the pairwise joint probability as
\begin{eqnarray}
\label{pm2}
 P(a_{1},b_{1} )=\frac{(1+ a_{1} \langle A_{1}\rangle+ b_{1} \langle B_{1}\rangle+ a_{1} b_{1} \langle A_{1}B_{1}\rangle)}{4}\nonumber\\
\end{eqnarray} 
and single marginals, for example, 
\begin{eqnarray}
\label{pms}
P(a_{1})=\frac{1+ a_{1} \langle A_{1}\rangle}{2} 
\end{eqnarray}
Similarly for $P(a_{1},a_{2},b_{2})$. Considering the non-negativity of any triple-wise joint probabilities and by adding four such suitable probabilities, we have 
\begin{eqnarray}
\label{c10}
&&P(a_{1},a_{2},-b_{1})+P(-a_{1},-a_{2},b_{1})+P(a_{1},-a_{2},-b_{2})\nonumber\\
&+&P(-a_{1},a_{2},b_{2})\geq0
\end{eqnarray}
A triple-wise joint probability, say, $P(a_{1},a_{2},-b_{1})$ can be written in terms of moment expansions is given by
\begin{eqnarray}
\label{c1}
&&P(a_{1},a_{2},-b_{1} )=(1/8)(1+ a_{1} \langle A_{1}\rangle+ a_{2} \langle A_{2}\rangle\nonumber\\
&-& b_{1} \langle B_{1}\rangle+ a_{1} a_{2} \langle A_{1}A_{2}\rangle- a_{2} b_{1} \langle A_{2}B_{1}\rangle\nonumber\\
&-&a_{1} b_{1} \langle A_{1}B_{1}\rangle- a_{1} a_{2} b_{1}\langle A_{1}A_{2}B_{1}\rangle )
\end{eqnarray}
 Putting Eq.(\ref{c1}) and three other such expressions in inequality (\ref{c10}), we have
\begin{eqnarray}
\label{c111}
CHSH&=&
a_{1}b_{2}\langle A_{1}B_{2}\rangle+a_{2}b_{1}\langle A_{2}B_{1}\rangle\\
\nonumber
&+& a_{1}b_{1}\langle A_{1}B_{1}\rangle-a_{2}b_{2}\langle A_{2}B_{2}\rangle-2\leq 0
\end{eqnarray}
which are CHSH inequalities. In quantum theory, for a suitable choice of state and observable it can be shown $(CHSH)_{Q}>0$, thereby violating the inequality (\ref{c111}).

 In the following, we show the equivalence between various formulations of CHSH inequalities using the above approach.
\subsection{Equivalence between various formulations of Bell's inequalities in local realist model and in quantum theory}
 Clauser and Horne \cite{ch} proposed a form of local realist inequalities in $2222$ scenario involving joint and single probabilities are the following
\begin{subequations}
\begin{eqnarray}
\nonumber
\label{chi1}
\nonumber
&&P(a_{1},b_{1})+P(a_{1},b_{2})-P(a_{2},b_{2})+P(a_{2},b_{1})\nonumber\\ 
&-&P(a_{1})-P(b_{1})\leq0\\
\nonumber
\label{chi2}
&&P(a_{1},b_{1})-P(a_{1},b_{2})+P(a_{2},b_{2})+P(a_{2},b_{1})\nonumber\\ 
&-&P(a_{2})-P(b_{1})\leq0\\
\nonumber
\label{chi3}
\nonumber
&&P(a_{1},b_{2})-P(a_{1},b_{1})+P(a_{2},b_{2})+P(a_{2},b_{1})\nonumber\\ 
&-&P(a_{2})-P(b_{2})\leq0\\
\nonumber
\label{chi4}
\nonumber
&&P(a_{1},b_{1})+P(a_{1},b_{2})+P(a_{2},b_{2})-P(a_{2},b_{1})\nonumber\\ 
&-&P(a_{1})-P(b_{2})\leq0\\
\nonumber
\end{eqnarray}
\end{subequations}
 which contains $64$ inequalities. 
Another interesting form termed as Wigner formulation \cite{ep} of local realistic inequalities in $2222$ scenario can be derived as
\begin{subequations}
\begin{eqnarray}
\nonumber\\
\label{bi11}
\nonumber
&&P(a_{1},b_{2})-P(a_{1},-b_{1})-P(a_{2},b_{2})\nonumber\\
&-&P(-a_{2},b_{1})\leq 0\\
\nonumber
\label{bi12}
&&P(a_{1},b_{1})-P(a_{1},b_{2})-P(a_{2},-b_{2})\nonumber\\
&-&P(-a_{2},b_{1})\leq 0\\
\nonumber
\label{bi13}
&&P(a_{2},b_{2})-P(a_{1},b_{1})-P(-a_{1},b_{2})\nonumber\\
&-&P(a_{2},-b_{1})\leq 0\\
\nonumber
\label{bi14}
\nonumber
&&P(a_{2},b_{1})-P(a_{1},b_{1})-P(a_{2},-b_{2})\nonumber\\
&-&P(-a_{1},b_{2})\leq 0\\
\nonumber
\end{eqnarray}
\end{subequations}
 which also contains $64$ inequalities. 
Using Eqs.(\ref{pm2}) and similar expressions, it can be shown that the above formulations of local realist inequalities given by Eqs.(\ref{chi1}-\ref{chi4}) and Eqs.(\ref{bi11}-\ref{bi14}) reduce to the CHSH inequalities in local realist models. 
The important question here is that whether such equivalence remains intact in quantum theory. In particular, we are interested if quantum expressions of left hand sides of the probabilistic inequalities (\ref{chi1}-\ref{chi4}) and (\ref{bi11}-\ref{bi14}) reduces to $(CHSH)_{Q}$ or not. In quantum theory, for a given state $\rho_{AB}$ the probabilities $P(a_{1},b_{1} )$ and $P(a_{1})$ can be written as
\begin{eqnarray}
\label{pmq2}
 P_{Q}(a_{1},b_{1} )=\frac{(1+ a_{1} \langle A_{1}\rangle_{Q}+ b_{1} \langle  B_{1}\rangle_{Q}+ a_{1} b_{1} \langle A_{1} B_{1}\rangle_{Q})}{4}\nonumber\\
\end{eqnarray} 
and single marginals, for example, 
\begin{eqnarray}
\label{pmqs}
P_{Q}(a_{1})=\frac{1+ a_{1} \langle A_{1} \rangle_{Q}}{2} 
\end{eqnarray}
Similar forms can be written for other probabilities. Note here the trivial fact again that no-signaling in space confirms that $\langle A_{1}\rangle_{Q}$ remains uninfluenced by the measurement of $B_{1}$ or $B_2$ and vice-versa. Putting Eqs.(\ref{pmq2} -\ref{pmqs}) and similar expressions in inequalities (\ref{chi1}-\ref{chi4}) and  inequalities (\ref{bi11}-\ref{bi14}), it is straightforward to check that the equivalence remains intact in quantum theory. In LG scenario, we shall see that this equivalence breaks down due the fact that the no-signaling in time condition is not in general satisfied in quantum theory. This allows us to find new sets of macrorealist inequalities inequivalent to standard LGIs. 
\section{Three-time Leggett-Garg scenario}
Let us consider the measurement of a dichotomic observable $\hat{M}$ having outcomes $\pm1$ is performed at time $t_1$,  $t_2$  and  $t_3$, which in turn can be considered as the measurement of the observables $\hat{M}_1$, $\hat{M}_2$, and  $\hat{M}_3$ respectively.
Then the measurement of the observables $\hat{M}_1$ , $\hat{M}_2$, and $\hat{M}_3$ should produce definite outcomes $+1$ or $-1$ at all instants of time from the assumptions of macrorealism \textit{per se}.
 Noninvasive measurability condition says that the outcomes of measurement of $\hat{M}_2$ or $\hat{M}_3$ remain unaffected due to measurement of $\hat{M}_1$ and so on. One can then formulate the standard LGIs is given by
\begin{eqnarray}
\label{lgi}
LG^{3}&=&m_{1}m_{2}\langle M_{1} M_{2}\rangle+ m_{2} m_{3}\langle M_{2}M_{3} \rangle\nonumber\\
&-&m_{1}m_{3}\langle M_{1}M_{3}\rangle-1\leq0
\end{eqnarray}
where $m_{1},m_{2},m_{3}=\pm1$. It is well studied that in quantum theory $(LG^{3})_{Q}>0$ for a suitable choice of observables, even for a qubit system. Note that the assumptions of macrorealism \textit{per se} and non-invasive mesurability imply the existence of joint probability distribution in a macrorealist model. 
\subsection{An alternate derivation of standard LGIs}
We provide an alternate derivation of standard LGIs by assuming the existence of joint probabilities $P(m_{1},m_{2},m_{3})$ of the outcomes $m_{1}$, $m_{2} $ and $m_{3}$. In a macrorealist model, the triple-wise joint probabilities $P(m_{1},-m_{2},m_{3})$ can be written as
\begin{eqnarray}
\label{c12}
&&P(m_{1},-m_{2},m_{3})=\frac{1}{8}[1+ m_{1} \langle M_{1}\rangle- m_{2} \langle M_{2}\rangle\nonumber\\
&+&m_{3}\langle M_{3}\rangle-m_{1} m_{2} \langle M_{1}M_{2}\rangle- m_{2} m_{3} \langle M_{2}M_{3}\rangle\nonumber\\
&+&m_{1} m_{3} \langle M_{1}M_{3}\rangle-m_{1} m_{2} m_{3}\langle M_{1}M_{2}M_{3}\rangle]
\end{eqnarray}
 Similarly for $P(-m_{1},m_{2},-m_{3})$. By choosing two such suitable triple-wise probabilities and by considering the positivity of probabilities we can write,
\begin{eqnarray}
\label{c11}
P(m_{1},-m_{2},m_{3} )+P(-m_{1},m_{2},-m_{3})\geq0
 \end{eqnarray}
Putting Eq.(\ref{c12}) and another such expression in the inequality (\ref{c11}), one gets 
the standard LGIs given by inequality (\ref{lgi}). Note that, only two triple-wise joint probabilities are enough to derive the standard LGIs. We now proceed to formulate various forms of LGIs and to examine their (in)equivalence with standard LGIs.
\subsection{A set of LGIs involving pair-wise anti-correlated probabilities}
 We first formulate a set of inequalities which are equivalent to standard LGIs both in macrorealistic theory and in quantum theory. For the derivation of this set of LGIs, the pair-wise probabilities only having anti-correlated outcomes are used. The existence of $P(m_{1},m_{2},m_{3})$ in macrorealist model enables one to calculate the appropriate  pair-wise marginals, for example, $P(m_{2},m_{3})=\sum_{m_{1}}P(m_{1},m_{2},m_{3})$. Similarly,  $P(m_{1},m_{2})$ and $P(m_{1},m_{3})$ can be obtained from appropriate marginalization. By  suitably choosing only the pair-wise probabilities of anti-correlated outcomes, we can write $\sum_{{m_{1}\neq m_{2}}}P(m_{1},m_{2})+\sum_{{m_{2}\neq m_{3}}}P(m_{2},m_{3})+\sum_{m_{1}\neq m_{3}}P(m_{1},m_{3})=2-2\left(P(m_{1},m_{2},m_{3})+P(-m_{1},-m_{2},-m_{3})\right)$. We can then formulate the following inequalities are given by
\begin{eqnarray}
\label{s1}
&&\sum_{m_{1}\neq m_{2}} P(m_{1}, m_{2})+\sum_{m_{2}\neq m_{3}} P(m_{2}, m_{3})\nonumber\\
&+&\sum_{m_{1}\neq m_{3}} P(m_{1},m_{3})\leq 2
\end{eqnarray}
Adopting similar approach, we obtain,
\begin{eqnarray}
\label{s2}
&&\sum_{{m_{1}\neq m_{2}}} P(m_{1},m_{2})-\sum_{m_{2}\neq m_{3}} P(m_{2},m_{3})\nonumber\\
&-&\sum_{m_{1}\neq m_{3}} P(m_{1},m_{3})\leq 0
\end{eqnarray}
\begin{eqnarray}
\label{s3}
&&\sum_{{m_{2}\neq m_{3}}} P(m_{2},m_{3})-\sum_{m_{1}\neq m_{2}} P(m_{1},m_{2})\nonumber\\
&-&\sum_{m_{1}\neq m_{3}} P(m_{1},m_{3})\leq 0
\end{eqnarray}
\begin{eqnarray}
\label{s4}
&&\sum_{m_{1}\neq m_{3}} P(m_{1},m_{3})-\sum_{{m_{2}\neq m_{3}}} P(m_{2},m_{3})\nonumber\\
&-&\sum_{m_{1}\neq m_{2}} P(m_{1},m_{2})\leq 0 
\end{eqnarray}
We shortly show that inequalities (\ref{s1})-(\ref{s4}) are equivalent to standard LGIs in a macrorealist model as well as in quantum theory, similar to CHSH scenario. This implies that whenever standard LGIs are satisfied, the inequalities(\ref{s1}-\ref{s4}) will also be satisfied. This inequalities are thus not useful for testing macrorealism over the standard LGIs. 
\subsection{Wigner and Clauser-Horne forms of LGIs}
We now formulate two different sets of inequalities belongs to the class (ii), i.e., the macrorealistic inequalities which are equivalent to standard LGIs in macrorealist theories but inequivalent in quantum theory. Note that, one set of such inequalities known as Wigner form of LGIs have already been proposed \cite{saha} and shown to be the stronger  than standard LGIs in quantum theory by us  \cite {pan17}. This was demonstrated by deriving disturbance inequalities from standard and Wigner form of LGIs. In this paper, we reproduced that argument in \cite{pan17} by using the approach mentioned earlier. 
 
Again from the assumptions of joint probability and non-invasive measurability, we obtain the pairwise statistics of measurement of $\hat{M_2}$ and  $\hat{M_3}$ having outcome $m_{1}$ and $m_{2}$ as 
$P(m_{2},m_{3})=\sum_{m_{1}=\pm}P(m_{1},m_{2},m_{3})$ and similarly for others. We can write the expression,
$P(-m_{1},m_{2})+P(m_{1},m_{3})-P(m_{2},m_{3})=P(-m_{1},m_{2},-m_{3})+P(m_{1},-m_{2},m_{3})$. By invoking the non-negativity of the probability,  Wigner form of LGIs \cite{saha, swati17, pan17} can be obtained as
\begin{eqnarray}
\label{w1}
P(m_{2},m_{3})-P(-m_{1},m_{2})-P(m_{1},m_{3})\leq 0		
\end{eqnarray}
One can obtain eight Wigner form of LGIs from (\ref{w1}).
Similarly, we derive 16 more inequalities are given by
\begin{eqnarray}
\label{w2}
P(m_{1},m_{3})-P(m_{1},-m_{2})-P(m_{2},m_{3})\leq 0
\end{eqnarray}
\begin{eqnarray}
\label{w3}
P(m_{1},m_{2})-P(m_{2},-m_{3})-P(m_{1},m_{3})\leq 0
\end{eqnarray}
Adopting the approach developed in Sec. II, we shall shortly show that Wigner form of LGIs are stronger than standard LGIs in quantum theory. 

Let us now propose a new form set of inequalities in class (ii), which we term as Clauser-Horne inequalities due to the presence of single probabilitis along with the pair-wise probabilities. The single marginal statistics of the measurement of the observable, for example, probability of getting outcome, when $M_{2}$ measurement is performed can be obtained as
$P(m_{2})=\sum_{m_{1},m_{3}=\pm}P(m_{1},m_{2},m_{3})$ and similarly for $P(m_{1})$ and  $P(m_{3})$. By combining single and pair-wise statistics, we can get the expression,
$P(m_{1},m_{3})+P(m_{2})-P(m_{1},m_{2})-P(m_{2},m_{3})=P(m_{1},-m_{2},m_{3})+P(-m_{1},m_{2},-m_{3})$, which in turn provides
\begin{eqnarray}
\label{ch1}
P(m_{1},m_{2})+P(m_{2},m_{3})-P(m_{1},m_{3})-P(m_{2})\leq0\nonumber\\
\end{eqnarray} 
 Similarly, $16$ more inequalities can be derived in this manner.
In compact notation, we can write,
\begin{eqnarray}
\label{ch2}
P(m_{1},m_{3})+P(m_{1},m_{2})-P(m_{2},m_{3})-P(m_{1})\leq0\nonumber\\
\end{eqnarray}
\begin{eqnarray}
\label{ch3}
P(m_{1},m_{3})+P(m_{2},m_{3})-P(m_{1},m_{2})-P(m_{3})\leq0\nonumber\\
\end{eqnarray}
The Clauser-Horne forms of LGIs given by Eqs.(\ref{ch1}-\ref{ch3}) can also be shown to be equivalent to standard LGIs in macrorealist model, but inequivalent to standard LGIs in quantum theory. Moreover, they will be shown to be stronger than Wigner form of LGIs in some specific cases.  
\subsection{Equivalence between  pair-wise anti-correlated form of LGIs and standard LGIs}
In order to examine the possible (in)equivalence among various formulations of LGIs, we write the pair-wise joint probability, for example, $P(m_{2},m_{3})$ in the moment expansion is given by
\begin{equation}
\label{i1}
P(m_{2},m_{3})=\frac{(1+ m_{2} \langle M_{2}\rangle+ m_{3}\langle M_{3}\rangle+m_{2}m_{3} \langle M_{2}M_{3}\rangle)}{4}
\end{equation}
Similarly, the single probabilities, for example, $P(m_{3})$ can be written as 
\begin{equation}
\label{i2}
P(m_{3})=\frac{(1+ m_{3}\langle M_{3}\rangle)}{2}
\end{equation}
 where $P(m_{3})=\sum_{m_{1},m_{2}=\pm} P(m_{1},m_{2},m_{3})$.

Putting the relevant pair-wise joint probabilities (as in Eq. (\ref{i1})) into that left hand side of the  Eqs.(\ref{s1}-\ref{s4}), one can obtain the standard LGIs given by Eq.(\ref{lgi}). Thus, all the pair-wise anti-correlated forms of LGIs are equivalent to standard LGIs in a macrorealisic theory. We examine whether such equivalence remains intact in quantum theory too.   

Given a density matrix $\rho$, in quantum theory a pair-wise sequential probability \cite{hall14} can be written as
\begin{eqnarray}
\label{p1}
P_{Q}(m_{1},m_{2})&=&\frac{1}{4}(1+ m_{1} \langle M_{1}\rangle_{Q}+ m_{2}\langle M_{2}^{(1)}\rangle_{Q}\nonumber\\
&+&m_{1} m_{2} \langle M_{1}M_{2}\rangle_{Q})
\end{eqnarray}
and a single probability is given by
\begin{equation}
\label{p2}
P_{Q}(m_{1})=\frac{(1+ m_{1}\langle M_{1}\rangle_{Q})}{2}
\end{equation}
 where the superscript in $\langle M_{2}^{(1)}\rangle_{Q}$ denotes that the measurement of $M_2$ in quantum theory is disturbed by the prior measurement $M_{1}$. Putting the expressions of joint probabilities similar to the one given by Eq.(\ref{p1}) in  the left hand side of the inequalities (\ref{s1})- (\ref{s4}), we get 
\begin{eqnarray}
\label{lgi1}
(LG^{3})_{Q}&=&m_{1}m_{2}{\langle M_1 M_2\rangle}_{Q}+m_{2}m_{3}{\langle M_2 M_3\rangle}_{Q}\nonumber\\
&-&m_{1}m_{3}{\langle M_1 M_3\rangle}_{Q}-1 
\end{eqnarray}
Hence, in quantum theory the left hand sides of the inequalities (\ref{s1})- (\ref{s4}) reduce to the quantum expression of LGIs. In other words, whenever a standard LGI is violated in quantum theory one obtains the violation of one of the inequalities (\ref{s1})- (\ref{s4}). Thus, such set of inequalities does not provide anything more than that is known from standard LGIs. This specific scenario is analogous to the Bell-CHSH scenario, where probabilistic formulations of local-realistic inequalities are equivalent to the CHSH one. However, we shall see that this is \textit{not} the case for Wigner and Clauser-Horne forms of LGIs which are inequivalent to and stronger than standard LGIs in quantum theory.

\subsection{Inequivalence between Wigner, Clauser-Horne and standard form of LGIs in quantum theory}
By using pairwise and single marginals, given by  Eq.(\ref{i1}) and Eq.(\ref{i2}) any of the $48$ Wigner and Clauser-Horne form of LGIs given by (\ref{w1})- (\ref{ch3}) reduces to one of the standard LGIs given by Eq.(\ref{lgi}). Hence, both Wigner and Clauser-Horne forms of LGIs are equivalent to standard LGIs in a macrorealist model. 
 
In quantum theory, by using Eq.(\ref{p1}) and similar quantities, the left hand side of $24$ Wigner form of LGIs in (\ref{w1})-(\ref{w3})  can be written  as
\begin{eqnarray}
\label{t1}
(W^{3})_{Q}&=&|\langle M_{2} \rangle_{Q}-\langle M_{2}^{(1)}\rangle_{Q}|\\
\nonumber
&+& |\langle M_{3}^{(2)}\rangle_{Q}-\langle M_{3}^{(1)}\rangle_{Q}|+(LG^{3})_{Q}
\end{eqnarray}
where $(LG^{3})_{Q}$ quantum expression of $LGI$ given by Eq.(\ref{lgi1}). If the measurement of $M_{1}$ does not disturb the statistics of $M_2$, then $\langle M_{2} \rangle_{Q}=\langle M_{2}^{(1)}\rangle_{Q}$ and if prior measurements do not disturb the statistics of $M_3$, so that, $\langle M_{3}^{(2)}\rangle_{Q}=\langle M_{3}^{(1)}\rangle_{Q}=\langle M_{3}\rangle_{Q}$. In that situation, Eq.(\ref{t1}) reduces to $(LG^{3})_{Q}$ only and we can say Wigner form of LGIs are equivalent to the standards ones in quantum theory. But in quantum theory,  $|\langle M_{2} \rangle_{Q}-\langle M_{2}^{(1)}\rangle_{Q}|\neq0$ and $|\langle M_{3}^{(2)} \rangle_{Q}-\langle M_{3}^{(1)}\rangle_{Q}|\neq0$, in general. Hence, from Eq. (\ref{t1}), we can say that the violation of standard LGIs implies the violation of Wigner form of LGIs, but the converse is not true. 

Hence, Wigner form of LGIs are stronger than the standard LGIs and captures the notion of macrorealism better than standard LGIs. We would like mention here that this inference regarding the inequivalence between Wigner and standard LGIs in quantum theory was proved earlier by us \cite{swati17}. But the approach adopted here is simple and elegant than the previous one.

Next, corresponding to $16$ Clauser-Horne form of LGIs given by inequalities (\ref{ch1})-(\ref{ch2}), using Eqs.(\ref{p1})-(\ref{p2}), we get similar form of Eq. (\ref{t1}). But, by using Eq.(\ref{p1})-(\ref{p2}) the left hand side of the $8$ inequalities given by (\ref{ch3}),  we get, 
\begin{eqnarray}
\label{t2}
(CH^{3})_{Q}&=&|\langle M_{2} \rangle-\langle M_{2}^{(1)}\rangle|+ |\langle M_{3}\rangle-\langle M_{3}^{(1)}\rangle|\nonumber\\
&+&|\langle M_{3}\rangle-\langle M_{3}^{(2)}\rangle|+(LG^{3})_{Q}
\end{eqnarray}
 Hence, following the above argument, we conclude that the Clauser-Horne form of LGIs are also stronger than the standard LGIs.

 Let us now compare Wigner form and Clauser-Horne form of LGIs. Eq. (\ref{t1}) contains only two additional terms apart from $(LG)_{Q}$, but Eq. (\ref{t2}) contains three  additional terms. Then the Clauser-Horne form of LGIs are inequivalent to the Wigner form of LGIs in quantum theory. From Eq.(\ref{t1}) and Eq.(\ref{t2}), we can write down the diffrence between Wigner and Clauser-Horne form as $(W^{3})_{Q}-(CH^{3})_{Q}=2(\langle M_{3}^{(2)}\rangle_{Q}-\langle M_{3}\rangle_{Q})=(1/2)\langle [M_{2}, M_{3}] M_{2}\rangle$, since $\langle M_{3}^{(2)}\rangle_{Q}=\langle M_{3}\rangle_{Q}+(1/2)\langle [M_{2}, M_{3}] M_{2}\rangle_{Q}$. Clearly, when $\langle [M_{2}, M_{3}] M_{2}\rangle_{Q}=0$, then the Clauser-Horne inequalities are equivalent to the Wigner form of LGIs. When $\langle [M_{2}, M_{3}] M_{2}\rangle_{Q}>0$, then the Wigner form of LGIs are stronger than the Clauser-Horne form of LGIs and if $\langle [M_{2}, M_{3}] M_{2}\rangle_{Q}<0$, then the Clauser-Horne form of LGIs are stronger than the Wigner form of LGIs. Therefore, which one is stronger than the other depends on the choice of measurement settings and states. However, both the probabilistic formulation of LGIs are stronger than the standard LGIs. This result has a direct relevance to experimentally testing the macrorealism through the LGIs. We can thus claim that the Wigner and Clauser-Horne forms of LGIs are better candidate than standard LGIs for testing macrorealism in quantum theory.
\section{Four-time Leggett-Garg scenario}
We extend the three-time LG scenario to four-time case and discuss the equivalent CHSH inequalities and inequivalen standard LGIs.
\subsection{Wigner form of LGIs are stronger than standard LGIs in quantum theory}
 We start by considering the existence of global joint probability distribution $P(m_{1},m_{2},m_{3},m_{4})$ for the measurements of $M_{1},M_{2},M_{3}$ and $M_{4}$ at $t_{1},t_{2},t_{3}$ and $t_{4}$ respectively. By suitable marginalization, various pair-wise and single  probabilities can be obtained. Now, invoking the non-negativity of probability following Wigner form of LGIs can be derived. In a compact manner, those can be written as 
\begin{subequations}
\begin{eqnarray}
\label{wlgic1}
&&P(m_{1},m_{2})-P(m_{1},m_{4})-P(m_{2},-m_{3})\nonumber\\
&-&P(m_{3},-m_{4})\leq 0
\\
\nonumber\\
\label{wlgic2}
\nonumber
&&P(m_{1},m_{4})-P(m_{1},m_{2})-P(-m_{2},m_{3})\nonumber\\
&-&P(-m_{3},m_{4})\leq 0
\\
\nonumber
\label{wlgic3}
\nonumber
&&P(m_{2},m_{3})-P(m_{1},m_{4})-P(-m_{1},m_{2})\nonumber\\
&-&P(m_{3},-m_{4})\leq 0
\\
\nonumber
\label{wlgic4}
\nonumber
&&P(m_{3},m_{4})-P(m_{1},m_{4})-P(-m_{1},m_{2})\nonumber\\
&-&P(-m_{2},m_{3})\leq 0
\end{eqnarray}
\end{subequations}
  It can be shown that the Wigner form of LGIs imply standard LGIs but the converse is not true in four-time measurement scenario too. By putting the sequential probabilities in quantum theory (as given in  Eq.(\ref{p1})) in the left hand side of the Wigner form of LGIs given by (\ref{wlgic1})-(\ref{wlgic4}), we get 
\begin{eqnarray}
\label{t5}
&&(W^{4})_{Q}=|\langle M_{2}\rangle_{Q}-\langle M_{2}^{(1)}\rangle_{Q}|+ |\langle M_{3}\rangle_{Q}-\langle M_{3}^{(2)}\rangle_{Q}|\nonumber\\
&+&|\langle M_{4}^{(1)}\rangle_{Q}-\langle M_{4}^{(3)}\rangle_{Q}|+ (LGI^{4})_{Q}
\end{eqnarray}
 where $(LGI^{4})_{Q}= m_{1}m_{2} \langle M_{1} M_{2}\rangle_{Q} +m_{2}m_{3} \langle M_{2} M_{3}\rangle_{Q} +m_{3} m_{4}\langle M_{3} M_{4}\rangle_{Q} -m_{1} m_{4}\langle M_{1} M_{4}\rangle_{Q}-2$.

In quantum theory, in general, $|\langle M_{2}\rangle_{Q}-\langle M_{2}^{(1)}\rangle_{Q}|\neq0$, $|\langle M_{3}\rangle_{Q}-\langle M_{3}^{(2)}\rangle_{Q}|\neq0$ and  $|\langle M_{4} ^{(1)}\rangle_{Q}-\langle M_{4}^{(3)}\rangle_{Q}|\neq0$. One can then conclude form Eq. (\ref{t5}) that Wigner form of LGIs in four-time measurement scenario are not only inequivalent but also stronger than standard four-time LGIs. 
\subsection{Clauser-Horne form of LGIs are stronger than the standard LGIs in quantum theory}
Following the similar procedure as adopted for three-time measurement, Clauser-Horne form of LGIs can also be obtained for four-time measurement scenario. We derive $64$ Clauser-Horne form of LGIs can compactly be written as 
\begin{subequations}
\begin{eqnarray}
\label{wlgch4}
\nonumber
&&P(m_{1},m_{2})-P(m_{1},m_{4})+P(m_{2},m_{3})\nonumber\\
&+&P(m_{3},m_{4})-P(m_{2})-P(m_{3})\leq 0
\\
\nonumber\\
\label{wlgch5}
\nonumber
&&P(m_{1},m_{2})+P(m_{1},m_{4})+P(m_{2},m_{3})\nonumber\\
&-&P(m_{3},m_{4})-P(m_{1})-P(m_{2})\leq 0
\\
\nonumber\\
\label{wlgch6}
\nonumber
&&P(m_{1},m_{2})+P(m_{1},m_{4})-P(m_{2},m_{3})\nonumber\\
&+&P(m_{3},m_{4})-P(m_{1})-P(m_{4})\leq 0
\\
\nonumber\\
\label{wlgch7}
\nonumber
&&P(m_{1},m_{4})+P(m_{2},m_{3})-P(m_{1},m_{2})\nonumber\\
&+&P(m_{3},m_{4})-P(m_{3})-P(m_{4})\leq 0
\end{eqnarray}
\end{subequations}
In quantum theory, by putting the quantum probabilities given by Eqs.(\ref{p1})-(\ref{p2})  in the left hand side of the Clauser-Horne form of LGIs in (\ref{wlgch4})-(\ref{wlgch5}), we get  the quantity $(W^{4})_{Q}$ given by Eq.(\ref{t5}).  But, by repeating the same procedure from (\ref{wlgch6})-(\ref{wlgch7}), we get 
\begin{eqnarray}
\label{t14}
\nonumber
&&(CH^{4})_{Q}=|\langle M_{2} \rangle-\langle M_{2}^{(1)}\rangle|+ |\langle M_{3} \rangle-\langle M_{3}^{(2)}\rangle|+|\langle M_{4}\rangle \nonumber\\
&-&\langle M_{4}^{(1)}\rangle|+|\langle M_{4}\rangle-\langle M_{4}^{(3)}\rangle|+(LGI^{4})_{Q}
\end{eqnarray}
Hence, we can say that both the Wigner and Clauser-Horne forms of LGIs are also stronger than the standard LGIs in four-time measurement scenario. This feature is in contrast to the Bell-CHSH scenario discussed in Sec.II.
\section{A set of inequivalent LGIs in two-time measurement scenario}
Let us now consider the LGIs for two-time measurement scenario. Halliwell \cite{hall17} argued that two-time LGIs provide the NSC for a weaker form of macrorealism. We write down  twelve two-time LGIs are given by
\begin{subequations}
\begin{eqnarray}
\label{th1}
LG_{12}=m_{1}\langle M_{1}\rangle+m_{2}\langle M_{2}\rangle-m_{1}m_{2} \langle M_{1}M_{2}\rangle-1\leq0\nonumber\\\\
\label{th2}
LG_{23}=m_{2}\langle M_{2}\rangle+m_{3}\langle M_{3}\rangle-m_{2}m_{3} \langle M_{2}M_{3}\rangle-1\leq0\nonumber\\\\
\label{th3}
LG_{13}=m_{1}\langle M_{1}\rangle+m_{3}\langle M_{3}\rangle-m_{1}m_{3} \langle M_{1}M_{3}\rangle-1\leq0\nonumber\\
\end{eqnarray}
\end{subequations}
 It can be shown that the probabilistic version of inequalities in two-time LG scenario are also stronger than the standard two-time LGIs given by (\ref{th1}-\ref{th3}). In order to showing this we propose the twelve such inequalities are given by 
\begin{subequations}
\begin{eqnarray}
\label{th4}
 PLG_{12}=P(m_{1})-P(-m_{2})- P(m_{1}, m_{2})\leq0\\
\label{th5}
 PLG_{23}=P(m_{2})-P(-m_{3})- P(m_{2}, m_{3})\leq0\\
\label{th6}
 PLG_{13}=P(m_{1})-P(-m_{3})- P(m_{1}, m_{3})\leq0
\end{eqnarray}
\end{subequations}
Putting the relevant pair-wise and single quantum probabilities given by Eqs.(\ref{p1})-(\ref{p2}) in the left hand side of (\ref{th4})- (\ref{th6}), we get
\begin{subequations}
\begin{eqnarray}
\label{th7}
(PLG_{12})_{Q}=|\langle M_{2}\rangle-\langle M_{2}^{(1)}\rangle|+(LG_{12})_{Q} \\
\label{th8}
(PLG_{12})_{Q}=|\langle M_{3}\rangle-\langle M_{3}^{(2)}\rangle| +(LG_{23})_{Q} \\
\label{th9}
(PLG_{12})_{Q}=|\langle M_{3}\rangle-\langle M_{3}^{(1)}\rangle|+(LG_{13})_{Q}
\end{eqnarray}
\end{subequations}
where $(LG_{12})_{Q}=m_{1}\langle M_{1}\rangle_{Q}+m_{2}\langle M_{2}^{1}\rangle_{Q}-m_{1}m_{2} \langle M_{1}M_{2}\rangle_{Q}-1$ and similarly for $(LG_{23})_{Q}$ and $(LG_{13})_{Q}$.
Following the argument that, in general, $|\langle M_{2}\rangle-\langle M_{2}^{(1)}\rangle|\neq0$ in quantum theory, we can say that inequality (\ref{th4}) is stronger than the inequality (\ref{th1}) and similarly for others. Note that, inequalities (\ref{th4})-(\ref{th6}) and (\ref{th1})-(\ref{th3}) respectively are equivalent in the macrorealist model.
\section{Curious forms of macrorealist inequalities}
In this section, we derive two new inequalities belong to class (iii), i.e., macrorealistic inequalities in three-time measurement scenario inequivalent to all the other above mentioned form of LGIs in  macrorealist model and in quantum theory. Interestingly, such inequalities capture the quantum violation of macrorealism even in parameter ranges where Wigner and Clauser-Horne forms of LGIs do not reveal the incompatibility between macrorealism and quantum theory. In Ref.\cite{pan17}, we had re-examined the relation between various formulations of LGIs, NSIT conditions and macrorealism. We had shown that for a given evolution, Wigner form of LGIs provide the violation for all values of $\tau$, except for two instants i.e., for the two particular qubit states $\rho(t_0)=|+\rangle\langle +|$ and $\rho(t_0)=I/2$ for the value of $\tau=\pi/4$ (will be explained shortly). We have found that Clauser-Horne form of LGIs for three measurement scenario and Wigner and Clauser-Horne forms of LGIs for four-time measurement scenario also do not show the quantum violation for the above two instances. Our proposed macrorealistic inequalities captures the quantum violation for those two specific instances.  

One of such macrorealist inequality can be derived as 
\begin{eqnarray}
\label{seq21a}
V_{1}&=&P(m_{3}=-1)-\sum_{m_{1}=m_{3}}P(m_{1},m_{3})\nonumber\\
&-&P(m_{2}=1,m_{3}=-1)\leq0
\end{eqnarray}
 The proof of inequality (\ref{seq21a}) goes as follows.  Let us consider the set of correlations $P(m_{2},m_{3})\in[P(m_{2}=1,m_{3}=1),P(m_{2}=1,m_{3}=-1),P(m_{2}=-1, m_{3}=1),P(m_{2}=-1,m_{3}=-1)]$ having four elements each. Similarly for $P(m_{1},m_{2})$ and  $P(m_{1},m_{3})$ where $m_{1},m_{2},m_{3}=\pm$. From the assumptions of non-invasive measurability, measurement of observable $\hat{M_{2}}$ does not disturb by the prior measurement $\hat{M_{1}}$ and similarly for $\hat{M_{3}}$. Also, future measurements do not disturb the prior measurement. Hence, we can write,
\begin{eqnarray}
\label{eq6}
P(m_{2})&\equiv&\sum_{m_{3}}P(m_{2},m_{3})=\sum_{m_{1}}P(m_{1},m_{2})
\end{eqnarray}
\begin{eqnarray}
\label{eq8}
P(m_{3})&\equiv&\sum_{m_{1}}P(m_{1},m_{3})=\sum_{m_{2}}P(m_{2},m_{3})
\end{eqnarray}
and
\begin{eqnarray}
\label{eq9}
P(m_{1})&\equiv&\sum_{m_{2}}P(m_{1},m_{2})=\sum_{m_{3}}P(m_{1},m_{3})
\end{eqnarray}
For notational convenience, we set $\gamma_{1}=P(m_{1}=1)$, $\gamma_{2}=P(m_{2}=1)$, $\gamma_{3}=P(m_{3}=1)$ 
and $\gamma_{12}=P(m_{1}=1,m_{2}=-1)+P(m_{1}=-1,m_{2}=1)$, $\gamma_{13}=P(m_{1}=1,m_{3}=-1)+P(m_{1}=-1,m_{3}=1)$ and 
$\gamma_{23}=P(m_{2}=1,m_{3}=-1)+P(m_{2}=-1,m_{3}=1)$. 
We can write,
\begin{align*}
P(m_{2}=1,m_{3}=-1)=(\gamma_{23}+\gamma_{2}-\gamma_{3})/2;
\end{align*}
\begin{align*}
P(m_{2}=-1,m_{3}=1)=(\gamma_{23}-\gamma_{2}+\gamma_{3})/2;
\end{align*}
\begin{align*}
P(m_{2}=1,m_{3}=1)=(-\gamma_{23}+\gamma_{2}+\gamma_{3})/2;
\end{align*}
\begin{align*}
 P(m_{2}=-1,m_{3}=-1)=1-(\gamma_{23}+\gamma_{2}+\gamma_{3})/2.
\end{align*}
\begin{figure}[h]
{\includegraphics[width=9cm]{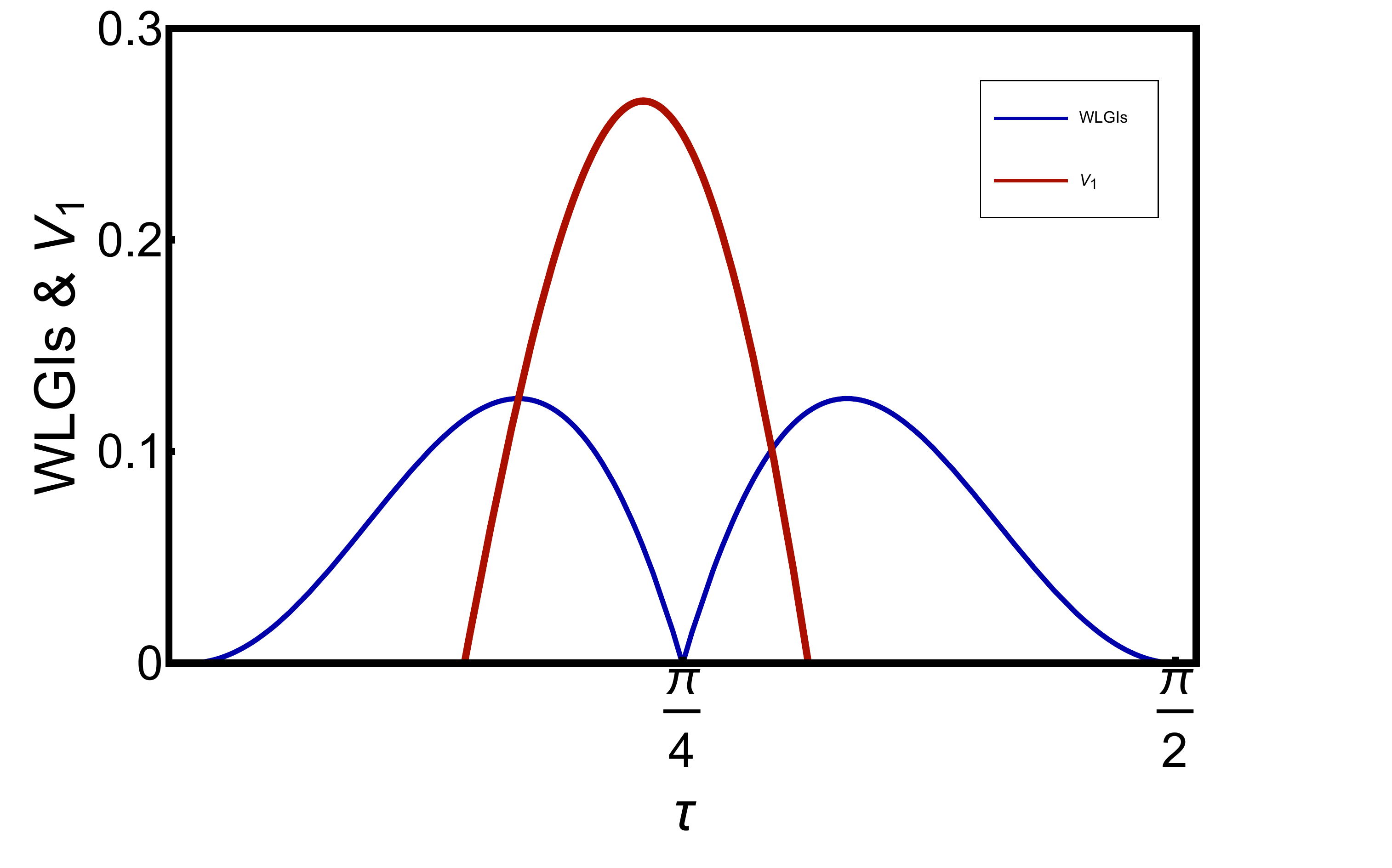}}
\caption{(Color online) The left had sides of the Wigner form of LGIs (\ref{w1}-\ref{w3}) and inequality (\ref{seq21a}) are plotted with respect to $\tau$ for the state $\rho=|+\rangle\langle +|$. It is seen that the quantum violation is obtained for \textit{any} value of $\tau$. The taller curve corresponds to the violation of inequality (\ref{seq21a}) and side curves corresponds the same for the violation of Wigner form of LGIs in (\ref{w1}-\ref{w3}). Similar curves can be found for $\rho=\mathbb{I}/2$.} 
\end{figure}
Similarly, $P(m_{1},m_{2})$ and $P(m_{1},m_{3})$ can be obtained for various combinations of $m_{1},m_{2},m_{3}=\pm1$.
Positivity condition of the probabilities $P(m_{2}=1,m_{3}=-1)$, $P(m_{2}=-1,m_{3}=1)$, $P(m_{2}=1,m_{3}=1)$ and $P(m_{2}=-1,m_{3}=-1)$ provide the condition, 
\begin{eqnarray}
\label{eq11}
|\gamma_{2}-\gamma_{3}|\leq \gamma_{23}\leq \gamma_{2}+\gamma_{3} \leq 2-\gamma_{23}
\end{eqnarray}
Using  the constraint (\ref{eq11}), we can write,
\begin{eqnarray}
\label{seq22a}
V_{1}&=&(1-\gamma_{3})-(1-\gamma_{13})-(-\gamma_{3}+\gamma_{2}+\gamma_{23})/2\nonumber\\
&=& (2\gamma_{13}-\gamma_{23}-\gamma_{3}-\gamma_{2}) /2\nonumber\\
&\leq& \big(2\gamma_{13}-\gamma_{23}-(2-\gamma_{23})\big)/2 \nonumber\\
&=& \gamma_{13}-1\leq0\nonumber
\end{eqnarray}
Similarly, using the constraint (\ref{eq11}), another inequality can be derived is given by
\begin{eqnarray}
\label{eq22a}
V_{2}&=&P(m_{2}=-1)-\sum_{m_{1}=m_{3}}P(m_{1},m_{3})\nonumber\\
&-&P(m_{2}=-1,m_{3}=1)\leq0
\end{eqnarray}

In order to demonstrate the usefulness of the inequality(\ref{seq21a}) and (\ref{seq22a}), let us consider a qubit state 
\begin{equation}
\label{qubit}
|{\psi(t_1)}\rangle = cos \theta |{0}\rangle +  \exp(-i\phi) sin \theta |{1}\rangle\nonumber
\end{equation} 
with $\theta \in [0,\pi]$ and $\phi \in [0,2\pi]$. Her $|{0}\rangle $ and $|{1}\rangle$ are the eigenfunctions of Pauli operator $\sigma_z$ having eigenvalues $\pm1$ respectively. The measurement observable at initial time $t_{1}$ is chosen to be  $M_{1}=\hat{\sigma_{z}}$ and the unitary evolution  $U_{ij} = \exp^{-i \omega (t_{j}-t_{i})\sigma_x}$, where $\omega$ is coupling constant and $\tau=|t_{j}-t_{i}|$ with $j>i$ and $i,j=1,2,3$. In \cite{pan17}, we have shown that except for the instant $\tau=\pi/4$ and for the qubit state $\rho=|+\rangle\langle +|$ or $\rho=I/2$, one can get the violation of $24$ Wigner form of LGIs (in Eqs.(\ref{w1}-\ref{w3})) for all values of $\tau$. We have also examined that  the Clauser-Horne inequalities (in Eqs.(\ref{ch1}-\ref{ch3})) are also not violated at that two specific instances. But, our formulated inequalities (\ref{seq21a}) or (\ref{eq22a}) are violated at above two instances (see Fig. 1). However, if we change the generator of the evolution, there may not be the violation of the inequality (\ref{seq21a}) or (\ref{eq22a}). But in such  case, following our prescription, new set of inequalities can be formulated, which may reveal the quantum violation macrorealism for the instances even when the Wigner and Clauser-Horne forms of LGIs do not show violation. We note again that  the inequalities (\ref{seq21a}) or (\ref{eq22a}) are inequivalent to all formulation of LGIs, both in macrorealist theory and in quantum theory.

\section{NSIT conditions and various formulations of LGIs }
An alternative formulation of macrorealism in terms of no-signaling in time (NSIT) conditions was proposed by  Clemente and Kofler \cite{clemente}. Although NSIT condition is analogous to the no-signaling in space condition in Bell scenario but the violation of the former do not produce any inconsistency in contrast to the violation of later.  The violation of a NSIT condition can be extrapolated at the level of individual measured value which implies that the NIM condition is violated.  
	
A general two-time NSIT condition can be read as 
\begin{equation}
NSIT_{(i)j}:P(m_{j})=\sum_{m_{i}} P(m_{i},m_{j})
\end{equation}
 which means that the probability of obtaining a particular outcome of the measurement of $M_{j}$  is unaffected by the prior measurement  $M_{i}$.\\

 Similarly, three-time condition $NSIT_{(1)23}$ can be written as 
\begin{eqnarray}
NSIT_{(1)23}:P(m_{2},m_{3})&=&\sum_{m_{1}} P(m_{1},m_{2},m_{3})
\end{eqnarray}

which states that  the pair-wise joint probabilities $P(m_{2},M_{3})$ are unaffected by the prior measurement $\hat{M_1}$.  Another three-time NSIT condition $NSIT_{1(2)3}$ is given by
 \begin{eqnarray}
NSIT_{1(2)3}:P(M_{1},M_{3})=\sum_{M_{2}} P(M_{1},M_{2},M_{3}).
\end{eqnarray}

It is discussed in \cite{maroney,clemente, swati17} that the aforementioned three-time NSIT conditions are necessary for standard LGIs but they do not provide the sufficient condition.The satisfaction of standard LGIs do not gurantee the satisfaction of one or both the NSIT conditions. We have shown in \cite{pan17} that three-time NSIT conditions  do not provide NSC for  Wigner form of LGIs too. 

Clemente and Kofler \cite{clemente} argued that the standard LGIs do not provide NSC for macrorealism. A conjunction of suitably chosen two-time and three-time NSIT conditions provides the NSC for macrorealism so that 
\begin{eqnarray}
\label{nsits}
NSIT_{(2)3} \wedge NSIT_{(1)23} \wedge NSIT_{1(2)3} \Leftrightarrow MR
\end{eqnarray}

Let us analyze the claim made in \cite{clemente}. In Bell scenario, pair-wise measurement of compatible observables are performed, whereas in LG scenario, the sequential measurement of incompatible observables are performed.  In LG scenario, the pair-wise sequential probabilities are obtained from the appropriate marginalization of the triple-wise sequential  probabilities. This is exactly the noninvasive measurability assumption at the ontic level which means that the prior measurement will not change the ontic state of the system and its subsequent dynamics. However, such an assumption does not hold in quantum theory, in general.

For the case of invasive sequential measurements in quantum theory, the most general form of the triple-wise sequential probability is given by,
\begin{eqnarray}
\label{c121}
&&P_{Q}(m_{1},m_{2},m_{3} )=(1/8)(1+ m_{1} \langle M_{1}\rangle_{Q}+m_{2} \langle M_{2}^{(1)}\rangle_{Q}\nonumber\\
&+& m_{3} \langle M_{3}^{(12)}\rangle_{Q}+m_{1} m_{2} \langle M_{1}M_{2}\rangle_{Q}+ m_{2} m_{3} \langle M_{2}M_{3}^{(1)}\rangle_{Q}\nonumber\\
&+&m_{1} m_{3} \langle M_{1}M_{3}^{(2)}\rangle_{Q}+ m_{1} m_{2} m_{3}\langle M_{1}M_{2}M_{3}\rangle_{Q} )
\end{eqnarray}
where $\langle M_{3}^{(12)}\rangle_{Q}$ denotes that quantum expectation value of  the observable $ M_{3}$ by taking the effect of measurements of both $ M_{1}$ and $ M_{2}$  into account. Similar explanation holds good for other such terms. It can be seen that if all the NSIT conditions used in Eq.(\ref{nsits}) are satisfied, Eq.(\ref{c121}) reduces to classical triple-wise joint probability \cite{hall17} similar to Eq. (11). This in turn imply that combination of NSIT conditions as used in Eq.(\ref{nsits}) provides the NSC for macrorealism. In other words, violation of one of the NSIT conditions warrants the violation of macrorealism. 

However, there is a crucial point to note regarding the testability of the alternate formulation of macrorealism based on NSIT conditions. As also indicated earlier, this proof can be considered as a logical proof of macrorealism. In order to verify the NSIT condition  a set of strict equalities needs to be experimentally tested. Such a precision is extremely difficult to achieve in a real experiment. This feature is quite similar to the Bell's original proof of non-locality \cite{bell} and Kochen-Specker logical proof of contextuality \cite{ks,ker}. In the former case, a perfect anti-correlation and in the later case, perfect predictability of the different outcomes were required to be tested in experiments. Since the presence of the noise is inevitable in any real experiment, such a requirement cannot be fulfilled. In order to fix this issue,  the CHSH inequalities \cite{chsh} and suitable non-contextual inequalities \cite{cabello} were proposed and tested experimentally. Therefore, suitable macrorealistic inequalities are required to be derived for testing the status of macrorealism in quantum theory.

\section{Summary and Discussions}

In this paper, we derived several new forms of macrorealistic inequalities different from standard LGIs. Since standard LGIs do not provide the necessary and sufficient conditions for macrorealism, then there remains a scope of formulating stronger macrorealist inequalities which can test the incompatibility between the macrorealism and quantum theory in a situation when standard LGIs fail to do so. We proposed three different classes of macrorealistic inequalities. First, a class of inequalities in three-time measurement scenario is derived by using the probabilities having only pair-wise anti-correlated outcomes which are  equivalent to standard LGIs in both macrorealist model and in quantum theory.  We then derived a different class of inequalities in three-time and four-time measurement scenario, termed as Wigner and Clauser-Horne forms. They are shown to be equivalent to the standard LGIs in macrorealist model but inequivalent and stronger in quantum theory. It is also shown that for a some specific cases the Clauser-Horne form can be shown to be stronger than Wigner form. We also proposed Wigner form of LGIs for two-time measurement scenario and demonstrate the inequivalence of them with the standard two-time LGIs in quantum theory.   Further, we derived another class of macrorealist inequalities which are inquivalent to the above two classes  in both macrorealist model and quantum theory. We have shown that this class of macrorealist inequalities captures the quantum violation of macrorealism even in the parameter ranges where no other LGIs in three-time measurement scenario reveal the incompatibility between macrorealism and quantum theory. 

In order to demonstrate our results, following \cite{hall14} we first developed a simple and elegant approach to provide alternate derivation of CHSH inequalities by assuming only the existence four triple-wise joint probabilities. This enables us to show the equivalence between probabilistic formulations local realistic inequalities and CHSH inequalities. Further, we provided an alternate derivation of standard LGIs by assuming only two triple-wise joint probabilities. The aforementioned approach is simple and elegant which we used to demonstrate the inequivalence between various classes of macrorealistic inequalities  and standard LGIs. We have already discussed that the above feature of inequivalence is in contrast to the CHSH scenario which arises due to following fact. In CHSH scenario, the statistical version of locality condition, i.e.,  the no-signalling in space condition is always satisfied in any physical theory. But, in LG scenario, the sequential measurement of non-commuting observables is performed. The statistical version of non-invasive measurability assumption, i.e., no-signaling in time condition is not satisfied in quantum theory in general. Due to such disturbance caused by a prior measurement to future measurements the inequivalent LGIs are found. 

We note again that in contrast to the Fine's theorem for CHSH inequalities, no set of LGIs do not provide the NSC condition for macrorealism. However, a suitable combinations of NSIT conditions provide the same \cite{clemente16,clemente}. One may then doubt the usefulness of LGIs for testing macrorealism.  As also mentioned earlier, the alternate formulation of macrorealism based on NSIT conditions  can be considered as a logical proof of macrorealism which requires to test a set of equalities. Such a demand is similar to the tests of Bell's original proof of non-locality \cite{bell} and the logical proof of Kochen-Specker theorem \cite{ks,ker}. Any real experiment will inevitably have uncontrollable noise in the measurement and hence the desired precision required for faithfully testing NSIT conditions is nearly impossible to achieve. One thus requires suitable macrorealist inequalities for testing the notion of marorealism in quantum theory.
We can argue that the inequivalent  inequalities derived here are better candidates for experimentally testing  macrorealism compared to standard LGIs. 

Finally, we note that by following our approach there is scope of formulating more set of inequalities other than the ones  derived here which may capture the violation of macrorealism in quantum theory for any given measurement situation. A critical analysis  is also required to examine the possible subtleties involved in the notions of macrorealism captured in various classes of LGIs and other macrorealist inequalities. Studies along the above mentioned lines could be an interesting avenue for future research.

\acknowledgments
 AKP acknowledges the support from Ramanujan Fellowship research grant (SB/S2/RJN-083/2014). 

\end{document}